\documentclass[11pt,fleqn]{article}
\usepackage{amsmath,graphicx}

\newcommand{\be}{\beq\label}
\newcommand{\ee}{\eeq}

\newcommand{\prt}{\partial}

\def\B{{\mathbb B}}
\def\C{{\mathbb C}}

\usepackage{amsfonts,amssymb,cite}
\usepackage{epsf,graphicx}

\def\noi{\noindent}

\newcommand{\Title}[1]{\noi {{\Large\bf #1}}\\[1ex]}

\def\Aunames#1{\noi{\bf #1}}
\def\auth#1{${}^{#1}$}
\def\Addresses#1{\medskip\noi \protect
	\begin{description}\itemsep -3pt {\it #1} \end{description}}
\def\addr#1#2{\item[${}^{#1}$]{\it #2}}

\newcommand{\Abstract}[1]{\vskip 2mm \begin{center}
        \parbox{16.4cm}{\small\noi #1} \end{center}\medskip}
\newcommand{\PACS}[1]{\begin{center}{\small PACS: #1}\end{center}}

\def\email#1#2{\footnotetext[#1]{e-mail: #2}\addtocounter{footnote}{1}}

\def\nqq{\hspace*{-2em}}

\def\Jl#1#2{#1 {\bf #2},\ }

\def\ApJ#1 {\Jl{Astroph. J.}{#1}}
\def\CQG#1 {\Jl{Class. Quantum Grav.}{#1}}
\def\DAN#1 {\Jl{Dokl. AN SSSR}{#1}}
\def\GC#1 {\Jl{Grav. Cosmol.}{#1}}
\def\GRG#1 {\Jl{Gen. Rel. Grav.}{#1}}
\def\JETF#1 {\Jl{Zh. Eksp. Teor. Fiz.}{#1}}
\def\JETP#1 {\Jl{Sov. Phys. JETP}{#1}}
\def\JHEP#1 {\Jl{JHEP}{#1}}
\def\JMP#1 {\Jl{J. Math. Phys.}{#1}}
\def\NPB#1 {\Jl{Nucl. Phys. B}{#1}}
\def\NP#1 {\Jl{Nucl. Phys.}{#1}}
\def\PLA#1 {\Jl{Phys. Lett. A}{#1}}
\def\PLB#1 {\Jl{Phys. Lett. B}{#1}}
\def\PRD#1 {\Jl{Phys. Rev. D}{#1}}
\def\PRL#1 {\Jl{Phys. Rev. Lett.}{#1}}

\def\lal{&&\nqq {}}

\def\beq{\begin{equation}}
\def\eeq{\end{equation}}
\def\bear{\begin{eqnarray}}
\def\bearr{\begin{eqnarray} \lal}
\def\ear{\end{eqnarray}}
\def\earn{\nonumber \end{eqnarray}}

\begin{document}
\twocolumn[

\Title{The algebrodynamics:  super-conservative collective dynamics \\ on  a ``Unique Worldline'' and the Hubble's law }

\Aunames{Vladimir V. Kassandrov\auth{a,1} and Ildus Sh. Khasanov \auth{b,2}}

\Addresses{
\addr a {Institute of Gravitation and Cosmology, Peoples' Friendship
	University of Russia,Moscow,117198 Russia}
\addr b {Scientific and Technological Center of Unique Instrumentation of Russian Academy of Sciences, Moscow, 117342 Russia}
	}

\Abstract
{We study the properties of roots of a polynomial system of equations which define a set of identical point particles located on a Unique Worldline (UW), in the spirit of the old Wheeler-Feynman's conception. As a consequence of the Vieta's formulas,  a great number of conservation laws is fulfilled for collective algebraic dynamics on the UW. These,  besides the canonical ones, include the laws with higher derivatives and those containing multi-particle correlation terms as well. On the other hand, such a ``super-conservative'' dynamics turns to be manifestly Lorentz invariant and quite nontrivial. At great values of ``cosmic time'' $t$ roots-particles demonstrate universal recession  (resembling that in the Milne's cosmology and simulating ``expansion'' of the Universe) for which the Hubble's law does hold true, with Hubble parameter being inversely proportional to $t$.}



] 
\email 1 {vkassan@sci.pfu.edu.ru}
\email 2 {khasanov@ntcup.ru}

\section{Implicit Unique Worldline, \\ ``one-electron Universe'' and \\ (bi)quaternionic analysis}  

We continue to explore the properties of correlated dynamics of point particles on a Unique Worldline (UW), in the spirit of the old Wheeler-Feynman' s conception~\cite{Feynman1,Feynman2}~\footnote{see also the works of E.C.G. Stueckelberg~\cite{Stueck}}.  In~\cite{JPhys1,Vest,JPhys2} it has been presented an algebraic realization of the UW concept. Specifically, let one has three  algebraic equations, 
\be{UW}
F_a(x,y,z,t) = 0,~( a=1,2,3),
\ee
with $\{F_a\}$ being three functionally independent {\it polynomials} dependent on 4 space-time (Cartesian) coordinates $\{x,y,z,t\}$. Then at any time moment $t=t_0$ the set of $N$ {\it roots} of system (\ref{UW}) fixes the positions of $N$ point particles which, in the course of time, demonstrate a correlated dynamics on the UW defined {\it implicitly} by (\ref{UW}). In fact, there are two kinds (R- and C-) of such particles corresponding to real (R-) or complex conjugate (C-) roots where the latter are represented in 3-space by their equal real parts. 

The power of such a realization of the Wheeler-Feynman UW program (which they also named ``one-electron Universe''~\footnote{``... why all electrons have the same charge and mass?''  --  ``Because they are  all the same electron!''~\cite{FeynmanPL}}) is that {\it for any nondegenerate polynomials the collective algebraic dynamics defined by (\ref{UW}) turns to be conservative.} Remarkably, conservation laws follow directly from the {\it Vieta's formulas} for polynomial equations~\cite{Vest}. For example, the first of these, 
\be{V1}
x_1+ x_2 + \cdots x_N = -a_1/a_0, 
\ee
where $\{x_i\},~i=1,2,\cdots N$ are the $x$-component of the roots of (\ref{UW}) and $a_0,a_1$ -- coefficients of the two leading monoms in corresponding polynomial equation $f(x,t)=0$ for $x$-component~\footnote{It results after elimination of the two other components $y,z$ from (\ref{UW})}. 

Now, for any (nondegenerate) form of the UW generating polynomials  the coefficient $a_0$ is a constant while $a_1$ is necessarily linear in $t$, see~Sect.4 for details. Thus, the formula (\ref{V1}) represents in fact {\it the law of inertial motion of the center of mass} for a system of $N$ particles of equal mass~\footnote{It follows from (\ref{V1}) that for any pair of complex conjugate roots the mass of corresponding C-particle should be twice greater than for R-particles}. 

Evidently, after differentiation by $t$ the law of conservation of ($x$-component) of total momentum does follow from (\ref{V1}). The (analogue of the) law of {\it energy conservation} follows from the second Vieta's formula.  Total {\it angular momentum} turns to be always constant in time, too. 

Now, we have considered the algebraic dynamics induced by (\ref{UW}) be {\it non-relativistic}, the time parameter $t$ playing the role of Newtonian absolute time. To pass to relativistic case, in~\cite{Chala} we have supplemented system (\ref{UW}) by the {\it equation of light cone}
\be{cone}
(T-t)^2 - (X-x)^2 - (Y-y)^2 -(Z-z)^2 =0, 
\ee
where the functions $X(T),Y(T),Z(T)$ dependent on the proper time $T$ specify the  worldline of a pointlike observer receiving light-like signals from the points-particles located on the UW. Corresponding system of equations (\ref{UW},\ref{cone}) is certainly {\it Lorentz invariant}, and  it happens that for any {\it inertially moving} observer the collective algebraic dynamics remains to be completely conservative~\footnote{Manifestly Lorentz invariant form of the conservation laws can be presented, see, e.g.~\cite{JPhys2}}. 
 
It is noteworthy that relativistic version of the UW concept can be realized in a simpler way~\cite{JPhys2} when the worldline is parameterized in a canonical explicit way, that is, through the functions $t(\tau),x(\tau),y(\tau),z(\tau)$ with monotonically varying timelike parameter $\tau$. 
Moreover, in the case of a time-like polynomially defined worldline one encounters with two remarkable properties of the roots-particles which take place at asymptotically great values of the proper time $T$ of the observer. Specifically, at some critical value $T=T_1$  the roots-particles start to couple while at  a greater moment of time $T=T_2>>T_1$ the formed pairs gather themselves into big groups -- clusters~\cite{JPhys2}. Unfortunately, for the case of implicitly defined UW the effects of pairing and clusterization had not been yet discovered.

Further, we want to pass to consider the fundamental origin of the algebraic dynamics, the so-called ``algebrodynamics''. It is based on the properties of (bi)quaternionic functions which resemble holomorphic functions of complex variable and are treated as primary physical fields. The original version of noncommutative analysis over the algebras of quaternion type (or, more widely, over matrix algebras) has been proposed long ago in our works (see, e.g., ~\cite{Q1,Q2,Q3} and references therein). Particularly, let one considers the algebra of {\it biquaternions} (complex quaternions) $\mathbb B$. Corresponding $\mathbb B$-differentiability conditions, natural generalization of the Cauchy-Riemann equations from complex analysis, {\it turn to be nonlinear as a direct consequence of non-commutativity of $\mathbb B$}. Thus, such ``good'' $\mathbb B$-functions can be naturally identified with fundamental {\it self-interacting} physical field.   

The principal equations of $\mathbb B$-differentiability are Lorentz invariant (on the Minkowski subspace of the vector space $\mathbb C^4$ of $\mathbb B$), possess natural spinor and gauge structures and can be reduced to a simpler form (the so called ``generating system of equations (GSE)) for effectively interacting 2-spinor and complex 4-vector fields. Moreover, general solution of the GSE can be obtained in an implicit algebraic form involving twistor structure. A whole set of free equations for canonical relativistic fields, in particular Maxwell equations, hold identically on the solutions of the GSE! As for particles, they can be identified with (isolated and bounded) singularities of the primary and secondary (gauge) $\mathbb B$-fields. 

Thus, {\it only from the $\mathbb B$-differentiability requirement}, one manages to deploy a wide picture of interacting fields/particles.  In many aspects, it  reproduces the canonical structures of relativistic physics.  On the other hand, in this framework it becomes possible to describe the collective dynamics in a simple and natural way. Just this is closely related to the Wheeler-Feynman's UW concept since any of the GSE solutions gives rise to  a {\it shear-free null congruence} which, on the other hand,  is naturally generated by point particles moving along a worldline.
            
The plan of the paper is as follows. In Sect.2, after a brief review of the principal features of $\B$-analysis and discussion on the $\B$-differentiability conditions and its reduced form (GSE), we  describe the general solution of GSE and its relation to the celebrated Kerr-Penrose   theorem and shear-free null congruences (SFC). In Sect.3, we demonstrate that for any UW, parameterized explicitly or implicitly, a principlal SFC can be constructed, and therefore defining equations of the UW are in fact Lorentz invariant if even the light cone equation (\ref{cone}) is not explicitly considered. 

In Sect.4, we describe the procedure of resolving system (\ref{UW}) of an implicitly defined UW for arbitrary nondegenerate polynomials $\{F_a\}$. By taking corrseponding {\it resultants}, one eliminates some two unknowns, say $y,z$, and analyzes the structure of the resulting polinomial equation $f(x,t)=0$, its Vieta's formulas and corresponding conservation laws. Then we formulate the concept of {\it super-conservative} dynamics which is not present in the canonical approaches and deals not only with additive constituents but with multi-particles' correlation terms. 

Finally, in Sect.5, we consider the properties of roots-particles of the UW equations (\ref{UW}) at great values of the time parameter $t$. We show that the roots demonstrate universal behavior, namely, recess from the center and each other. Such a process resembles that of the expansion of the Universe and, in particular, the ``kinematical'' cosmology of E.A. Milne~\cite{Milne}. Calculation of radial velocities leads to the conclusion that the Hubble's law does hold and that the Hubble parameter is inversely proportional to the value of cosmic time $t$.  
In the conclusion (Sect.6) we discuss some properties and perspectives of the UW approach. 

 \section{$\B$-analysis and generating \\system of equations}
 
Natural generalization of the complex analysis to the case of associative but noncommutative algebras had been proposed in our works (see, e.g.,~\cite{Q1,Q2,Q3} and references therein). Physical application is possible in the case of the {\it biquaternion} algebra $\B$ which is isomorphic to the full matrix algebra $Mat(2,\C)$.       
 Its automorphism group $SO(3,\mathbb C)$ is isomorphic, $SO(3,\mathbb C) \sim SL(2,\mathbb C)$, to the 6-parametric spinor Lorentz group $SL(2,\mathbb C)$. That's why $\mathbb B$-algebra looks as a natural candidate to be the {\it space-time algebra}!
    
 Specifically, one requires for a $\B$-valued function $F(Z)\in \B$ be {\it $\mathbb B$-differentiable} if the folowing relation does hold: 
 \be{kass}
 dF = \Phi \cdot dZ \cdot \Psi, 
 \ee
 where $\Phi=\Phi(Z),~\Psi=\Psi(Z)$ are two auxiliary $\mathbb B$-valued functions which can be named ``semi-derivatives'' (left and right, respectively). 
 
 Thus, {\it a $\mathbb B$-valued function is $\mathbb B$-differentiable if its differential $dF$ can be represented in a component-free form, that is, via the operation of multiplication $(\cdot)$ in $\mathbb B$}. It turns out that 
 the class of $\mathbb B$-differentiable functions is rather wide so that they can be treated as fundamental physical fields~\cite{Q2,YadPhys}.

 However, vector space of $\mathbb B$ is 8-dimensional in reals, so one has well-known problems with interpretation of the 4 superfluous coordinates. Moreover, the Minkowski space $\bf M$ does not even form a subalgebra of $\mathbb B$. Therefore, the first version of {\it algebrodynamics} had been successfully developed under the assumption that the coordinate space in (\ref{kass}) is restricted to the Hermitian matrices $Z \mapsto X=X^+$, that is to the Minkowski space $\bf M$. In the procedure, however, the fields $F(X),\Phi(X),\Psi(X)$ are still complex-valued as this is frequently accepted in physics. 
  
Besides, below we restrict ourselves by the simplest and most remarkable case when one of the ``semi-derivatives'', say, $\Psi=\Psi(X)$ coincides with the principal function $F(X)$. Thus, in the framework of algebrodynamics we consider the following reduced form of $\mathbb B$-differentiability conditions (\ref{kass}):
\be{GSEF}
dF = \Phi \cdot dX \cdot F,
\ee
where $X=X^+ = x^0 {\bf 1} + x^{a}  {\bf \sigma}_{a}$,~~$\{x^\mu\},~\mu=0,1,2,3$ are the Minkowski space-time coordinates, while $\bf 1$ and $\{{\bf \sigma}_{a}\}, ~a=1,2,3$ -- the $2\times 2$ unit matrix and three Pauli matrices, respectively. 
    
Since in the full $2\times 2$ matrix representation of $\mathbb B$ both columns of $F(X)$ are completely independent, one can compose any solution of (\ref{GSEF}) from a solution of the following its reduced form:
\be{GSES}
d\xi = \Phi dX \xi, 
\ee
where $\xi=\xi(X)$ is one of the columns of the matrix $F(X)$. Further on  system (\ref{GSES}) is called the {\it generating system of equations} (GSE).

Reduced conditions of $\mathbb B$-differentiability (\ref{GSEF}) or, equivalently, (\ref{GSES}) constitute a full set of equations of an {\it algebraic field theory}, the so-called ``algebrodynamics''~\cite{Q1,YadPhys,Vest0,Pavlov}. Such a theory is Lorentz invariant and contains a natural gauge and 2-spinor (twistor) structure (see below). It allows for definition of a set of fundamental relativistic fields whose free equations do hold identically on the solutions of (\ref{GSES}). Note that system (\ref{GSES}) is over-determined, so that both spinor $\xi(X)$ and vector $\Phi(X)$ fields can be found from it (see below). 

 As for particles, {\it one can identify them with (isolated and bounded) singularities} of the primordial fields $\xi(X),\Phi(X)$ or secondary gauge fields which can be defined via the latter. Singular locus can be represented by isolated points, strings or 2D membranes and demonstrate nontrivial time dynamics (see, e.g.,~\cite{Sing}.

\section{Twistor structure and general solution to the equations of  \\biquaternionic differentiability}

 We have seen that in the simplest and most remarkable  case equations of  $\B$-differentiability reduce to the following (GSE) form (\ref{GSES}):
 \be{GSEN}
 d\xi = \Phi dX \xi. 
 \ee
 Rewriting now system (\ref{GSEN}), one obtains:
 \be{link}
  d\xi=\Phi d(X\xi) - \Phi X d\xi, \rightarrow (I+\Phi X) d\xi = \Phi d\tau, 
  \ee
 where the 2-spinor 
 \be{incid2}
 \tau:= X\xi
 \ee
  together with the primary 2 -spinor $\xi$ forms a {\it twistor} on the Minkowski space-time $\bf M$~\footnote{We may ignore in the incidence relation (\ref{incid2}) the frequently used factor $i$} .    

Relation (\ref{link}) indicates that 4 twistor components, that is, the two 2-spinors $\{\xi,\tau\}$ are {\it functionally dependent}. Presisely, for any solution of (\ref{GSEN}) there exist two  independent implicit functions $\{\Pi^C\},~ C=1,2$ of four complex arguments such that one has 
\be{Gensol}
\Pi^C (\xi,\tau) = 0.
\ee
In fact, the two algebraic equations (\ref{Gensol}) represent {\it general solution} to the equations of the reduced form GSE of  $\B$-differentiability conditions. Indeed, at any point $X \in \bf M$ the two equations 
(\ref{Gensol}), that is,  
\be{GensolV}
\Pi^C (\xi, X\xi)= 0,
\ee
can be resolved w.r.t. the two unknowns $\{\xi^A\},~A=1,2$. Executing the procedure at every point $X$ and selecting a continious branch of the roots of (\ref{GensolV}), one obtains a spinor field $\xi(X)$. Generally, there exists a whole set of the continious branches of such a {\it milti-valued} field.

Making use of the projective-like invariance of   (\ref{GensolV}) under the transformation 
\be{proj}
\xi \mapsto \alpha(\xi,\tau)\xi,~~\tau\mapsto \alpha(\xi,\tau)\xi, 
\ee
one can reduce it to only one constraint defining, say, the ratio $G$ of components of the 2-spinor,  
\be{ratio}
G:=\xi^2/\xi^1. 
\ee
Then, choosing the projective chart as $\xi^1 \mapsto 1,~\xi^2  \mapsto G$,  one obtains, instead of (\ref{GensolV}), general solution to (\ref{GSEN}) in the following form:
\be{gensim}
\Pi (G, \tau_1,\tau_2)=0, ~~\tau_1=wG+u,~\tau_2 =vG+\bar w,
\ee
where $u,v = ct\pm z,~\bar w, w = x\pm iy$ are the spinor coordinates which compose the Hermitian matrix $X=X^+$ representing the points of $\bf M$,
\be{herm}
X =
\left(
\begin{array}{cc}
u & w \\
\bar w & v 
\end{array}
\right) . 
\ee
 
Now, the constraint (\ref{gensim}) is well-known to generate a null shear-free congruence (NSFC) of light-like rays on $\bf M$. According to the Kerr-Penrose theorem~\cite{Penrose}, all analytical NSFC can be constructed via resolving of an algebraic constraint isomorphic to (\ref{gensim}). In particular, differentiating the latter w.r.t. spinor coordinates, one obtains the equations
\be{SFC}
\prt_w G = G\prt_u G, ~~\prt_v G = G \prt_{\bar w} G,
\ee
which are just the defining equatons of a NSFC, with $G$ representing its principal spinor. It is easy to check using (\ref{SFC}) that $G$ satisfies both fundamental relativistic equations, the (complex) eikonal one,
\be{eik}
\vert {\underline \nabla} G \vert ^2:=\prt_u G \prt_v G -\prt_w G \prt_{\bar w} G =0, 
\ee 
and the linear wave equation,   
\be{wave}
\Box G :=\prt_u  \prt_v G -\prt_w  \prt_{\bar w} G =0, 
\ee 
respectively. 

Note that equations (\ref{SFC}) follow from (\ref{gensim}) everywhere except the points where the roots of the latter are mupliple, that is 
\be{sing}
\det \Vert \frac{d\Pi}{dG} \Vert = 0.
\ee 
This is the condition defining the {\it caustic} points of the NSFC at which the derivatives of $G$ become infinite. At the same time (\ref{sing}) specifies the {\it singular locus} of secondary gauge fields which can be defined through the matrix-valued field $\Phi$ in the GSE equations (\ref{GSEN}). In general framework of $\B$-algebrodynamics {\it particles are considered as singularities of these fields, so that (\ref{gensim}) together with (\ref{sing}) defines  their spacial distribution, at a fixed moment, and time dynamics as well~\cite{Sing}}. 

\section{Vieta's formulas and \\ super-conservative dynamics}     

It is well known that a particle moving along an arbitrary worldline $X=X_0(s)$ gives rise to a system of light-like rays which form just the NSFC. For such a congruence, {\it divergence} is the only nonzero optical invariant~\footnote{A worldline located on $\C\bf M$, complex generalization of $\bf M$, generates an NSFC with nonzero {\it torsion}, familiar example of such being represented by the Kerr congruence}. The null congruence can be immediately constructed from the requirement of equality of local twistor components for the point of observation $X$ and corresponding point on the worldline $X_0(s)$, 
\be{twist}
\tau = X_0(s)\xi =X\xi.
\ee
In components, (\ref{twistcomp}) gives rise to two equations, 
\be{twistcomp}
\begin{array}{cc}
\tau_1 - w_0(s)) G - u_0(s))=0, \\ 
\tau_2 - v_0(s))G - \bar w_0(s)) =0, 
\end{array}
\ee
from which, after elimination of the parameter $s$, one comes to the relation reproducing (\ref{gensim}), that is, to functional dependence of three projective twistor components $G,\tau_1,\tau_2$.  That means that the congruence defined at any point by (\ref{twist}) 
is necessarily null and shear-free. 
 
In fact, the same conclusion can be made for any implicitly defined (Unique) Worldline (UW) defined by (\ref{UW}). Rewriting the latter for convinience in spinor coordinates, 
 \be{UWS}
 F_a(u,v,w,\bar w)=0, ~(a=1,2,3),
 \ee
we supplement this system with the twistor definitions, analogous to (\ref{twistcomp}),
 \be{twistcomp2}
 \tau_1 - w G - u=0,~~\tau_2 - vG - \bar w =0,
 \ee
where {\it all of the coordinates $u,v,w,\bar w$ correspond to a point on the UW}, that is, not to the point of observation!  Eliminating then 4 coordinates from 3 equations (\ref{UWS}) together with 2 equations (\ref{twistcomp2}), 
one arrives again at the principal functional dependence of the form (\ref{gensim}), 
\be{gensim2}
\Pi(G,\tau_1,\tau_2)=0, 
\ee
affirming that the null congruence generated by the UW (\ref{UWS}) is again shear-free. As a consequence, {\it the implicit UW construction turns to be manifestly Lorentz invariant!}~\footnote{Remarkably, in preceding works~\cite{JPhys1}, \cite{JPhys2} we consider the UW construction based on (\ref{UW}) as non-relativistic, with $t$ playing the role of absolute Newtonian time}

Explicitly, under a transformation of coordinates, 
\be{lor}
X \mapsto S^+ X S,~~ S\in SL(2,\C),
\ee
with $SL(2,\C)$  being the spinor Lorentz group, corresponding components of the twistor maps as 
\be{twistmap}     
 \xi \mapsto S^{-1}\xi, ~~\tau \mapsto (S^+)^{-1}\tau, 
\ee
 so that  the UW equations (\ref{UWS}) together with principal twistor constraint (\ref{gensim2}) defining the NSFC remain form-invariant~\footnote{In the procedure, the ratio of spinor components $G$ transforms according to a linear-fractional law}. 

Let us now concentrate on the principal polynomial case of the UW equations (\ref{UW}). We assume that all three polynomials $\{F_a\}$ are functionally independent and nondegenerate. The first assumption results in a finite set of roots of (\ref{UW}) which, both real and complex conjugate, can all be found, at least numerically. Nondegenerate form of the polynomials means that all the terms of the leading monoms are nonzero. For example, in the case of quadratic polynomials one assumes 
\be{monoms}
\begin{array}{cc}
F_1 = \Sigma_{i+j+k+l=2}=2 a_{ijkl} x^i y^j z^k t^l +\\ 
+ monoms~linear~~ in~ x,y,z,t +~a~constant =0, 
\end{array}
\ee
$a_{ijkl} \ne 0$, and for $F_2,F_3$ analogously. 

To resolve system (\ref{UW}), one should eliminate some two unknowns, say, $y,z$ by successive computing corresponding {\it resultants}. Techniques  of this procedure is well known~\cite{Uteshev}, based on the structure of the so-called {\it Sylvester matrices} and, in particular, described in detail in \cite{Vest}. In result, we obtain a polynomial equation $f(x,t)=0$ whose leading monoms have the form
\be{monomsXT}
\begin{array}{ccc}
f(x,t) = \Sigma_{i+j=N} (c_{ik} x^i t^j) + \\ 
+~monoms~of~lower~orders~= \\
c_{00} x^N +c_{01} x^{N-1} t + \cdots + c_{NN} t^N  + \cdots =0,
\end{array}
\ee
where $c_{ij}$ are real constants and the degree $N=N_1 N_2 N_3$ is just the product of degrees $N_1,N_2,N_3$ of the three generating polynomials $\{F_a\}$. Making use of (\ref{monomsXT}), one can analyze the structure of the Vieta's formulas linking its roots. For this, let us rewrite the latter,  collecting all the terms  leading in $x$, 
\be{previeta}
f(x,t) = A_0 x^N + A_1(t) x^{N-1} + A_2(t) x^{N-2} + \cdots =0,
\ee
with $A_0=a_0,~A_1(t)= a_1 t + b_1,~A_2(t)= a_2 t^2 +b_2 t +c_2$ and 
$a_i,b_i,c_i$ being again some constant coefficients. 

Now, writing out the first Vieta's formula 
\be{vieta1}
x_1 + x_2 + \cdots + x_N = -\frac{A_1(t)}{A_0}=-\frac{a_1 t +b_1}{a_0}
\ee
and differentiating it w.r.t. $t$, one immegiately arrives at the law of conservation of $x$-projection of the total momentum for the system of $N$ roots-particles represented by (\ref{UW}), 
\be{Xmomentum}
\dot x_1 + \dot x_2 + \cdots +\dot x_N = -\frac{a_1}{a_0} = constant 
\ee      
(here the masses of all particles are equal and set unit). The same property, certainly, takes place for two other projections of the momentum. 
 
Consider now the second Vieta's formula, 
\be{vieta2}
\Sigma_{i\ne j} (x_i x_j) =\frac{A_2(t)}{A_0}=\frac{a_2 t^2 +b_2 t +c_2}{a_0},
\ee   
twice differentiating which, one obtains the following conservation law:
\be{sconserv}
\Sigma_{i\ne j} (\dot x_i \dot x_j + \frac{1}{2}x_i \ddot x_j) =\frac{ a_2}{a_0}= constant.
\ee
Structure of the conserved quantity in the l.h.p. of (\ref{sconserv}) is rather unusual and, to our knowledge, does not appear in the framework of canonical mechanics. It consists of two kinds of {\it correlation terms} one of which corresponds to {\it pairwise correlations} of momenta (velocities) of different particles while the second one relates the coordinate of each particle with the sum of accelerations of the others. Certainly, two  analogous relations can be obtained for $y,z$-components after elimination of corresponding unknowns. 

 It is also evident that for a system (\ref{UW}) with $N>2$ roots there exist a set of conservation laws with {\it higher order derivatives} linking correspondent characteristics of different roots-particles. Such an unusual dynamics can be called {\it super-conservative}. 
 
One should also bear in mind that the existence of correlation type conservation laws does not forbid the existence of usual additive conservation laws, including not only  the total momentum but the analogue of total energy and the laws with higher derivatives as well. For example, the second Vieta's formula (\ref{vieta2}), in account of the first linear one (\ref{vieta1}), can be rewrited {\it in the additive form},
\be{vieta2A}
\begin{array}{cc}
\Sigma_{i=1}^N (x_i)^2 =(\frac{A_1}{A_0})^2 - 2\frac{A_2}{A_0} = \\
(\frac{a_1 t +b1}{a_0})^2 - 2\frac{a_2 t^2 +b_2 t +c_2}{a_0}, 
 \end{array}
 \ee   
 from which, after two ditterentiations w.r.t. $t$, it follows:
 \be{virial}
 \Sigma_{i=1}^N  (\dot x_i^2 +x_i \ddot x_i) = \frac{a_1}{a_0} - 2\frac{a_2}{a_0} = const. 
 \ee
  Evidently, the same is valid for $y,z$-components, and summing the three relations up, one obtains a {\it $SO(3)$-invariant} conservation law,  
 \be{SO(3)}
\Sigma_{i=1}^N  (\dot {\bf r}_i^2 +{\bf r}_i  \ddot {\bf r_i}) = const, 
\ee
 where ${\bf r}_i,\dot {\bf r}_i, \ddot {\bf r}_i$ are the radius-vector, velocity and acceleration of the $i$-th particle, respectively. The first term in (\ref{SO(3)}) reproduces doubled kinetic energy, while the second one  relates again the position and  acceleration of (equal to the resultant force acting  on) the $i$-th particle and resembles thus the {\it virial term} in mechanics~\cite{Virial}. Thus, the conservation law resulting from the second Vieta's formula, represents an analogue of the law of energy conservation or, presisely, the (reinforced) virial theorem ~\cite{diss}!
 
Finally, the law of conservation of total {\it angular momentum} is also valid though it does not follow explicitly from the Vieta's formulas. The procedure of exact calculation of this quantity for a particular generating system (\ref{UW}) is described in detail in~\cite{JPhys2, Chala, diss}, together with numerous illustrative examples.    

To conclude, such a rigid system of conserved quantities, of both additive and correlation characters, does not lead to a trivial collective dynamics of roots-particles. On the contrary, though all the roots-particles belong to the sole UW, an external observer considers them as a system of interacting but {\it individual} particles involved in a complicated yet {\it super-conservative} collective dynamics. Some 2D and 3D examples are presented in \cite{Chala, diss}, respectively.

\section{Recession of roots and \\ the Hubble's law}

 Let us return now to analyse the polynomial equation (\ref{monomsXT}) obtained from the generating system (\ref{UW}) after elimination of two unknowns $y,z$. Consider the region of great values of the time parameter $t$, define the $x$-projection of the {\it velocity} of a root $u$,
 \be{veloc}
 x=u t, 
 \ee

and rearrange the monoms in (\ref{monomsXT}) collecting the terns of leading order in $t$. Then one gets
\be{great}
\begin{array}{cc}
(c_{00} u^N + c_{01} u^{N-1}+ \cdots + c_{NN}) t^N + \\ 
+~monoms~ of~lower~orders~ in~t = 0.
\end{array}
\ee  
From (\ref{great}) one immegiately finds that, at great values of $t$, velocity $u$ of any of $N$ roots approximates a constant while acceleration approaches zero. Evidently, for $y,z$-projections of velocities $v,w$ one can make again the same conclusion, so that the radial component $V$ of  velocity of any root 
\be{rad}
V=\frac{ux}{r}+ \frac{vy}{r}+\frac{wz}{r} = \frac{r}{t}
\ee
 is proportional to the distance $r$ of a root from the origin, $r=\sqrt{u^2+v^2+w^2} t$. 
 
 In result, one comes to the Hubble's law with the {\it Hubble parameter} $H$ being inversely proportional to the ``cosmic time'' $t$.  
\be{Hubble}
H=\frac{1}{t}
\ee
and universal for all of the roots-particles. The same is certainly valid for a pair of complex congugate roots whether one represents them as a unique particle (of double mass) w.r.t. their equal real parts. 

Thus, one encounters the effect of asymptotically uniform {\it expansion} of the ``roots-Universe'' with the most simple law of the variation of Hubble parameter completely compatible with modern observations and representations~\cite{Observ2}.

It is noteworthy to mention that the ``kinematical'' approach to ``cosmological'' evolution, together with the law (\ref{Hubble}), strongly resembles the old theory of E.A. Milne~\cite{Milne}. Remarkably, such a simple theory predicts a number of cosmological consequences quite close to those of the $\Lambda$CDM model~\cite{Observ, Silagad}. 

However, whether in the Milne's model  mutual recession (``expansion'') starts from a singular point, in our approach the evolution is infinite both in the future and the past. Specifically, one can consider the great {\it negative} values of the time parameter $t$ when the {\it contraction} of the ``roots-Universe'' takes place. When the values of $t$ decrease in modulus and approaches zero, one observes a complicated picture of individual dynamics of roots-particles. It includes, in particular, their numerous $R-C$ transmutations modelling the creation/annihilation processes, and increase of the ``effective temperature'' of the roots-particles' ensemble.  {\it None singularity does ever occure}

 \section{Conclusion}
 
  In the paper we have considered the most rich (in internal properties) and effective way to realize the Unique Worldline (UW) concept. Specifically, we analyzed the structure of equations for an implicitly defined and polynomially parameterized UW whose (real and complex congugate) roots correspond to a set of point particles of two kinds (R- and C-, respectively). They participate in a nontrivial temporal dynamics which turns to be {\it super-conservative}, that is, satisfying a large set of conservation laws of both additive and correlation nature, the latter being quite unknown in classical mechanics. Remarkably, all these (except the law of conservation of angular momentum) follow directly from the structure of the Vieta's formula for the roots of the UW equations. Contrary to the previous conviction, the whole procedure turns to be manifestly Lorentz invariant by itself, without any consideration of an external observer with his light cone (as it was earlier supposed in~\cite{JPhys2,Chala}. 
  
One of the important obtained results is the universal {\it asymptotic} behavior of roots-particles  at great values of ``cosmic time'' $t$. In fact, it resembles the process of recession of ``fundamental particles'' in the ``kinematical'' Milne's cosmology and results in the Hubble's law, with the Hubble parameter being universal and inversely proportional to $t$. However, in the past the faith of the ``roots-Universe'' strongly differs from the Milne's picture (which starts from a singular point of Big Bang) containing the stages of contraction and nontrivial nonsingular dynamics near $t=0$ and small distances from the origin. Unfortunately, the effects of pairing and successive aggregation specific for a timelike explicitly parameterized polynomial UW were not yet discovered for  an implicitly defined UW.  In future we are going to find a way to join the advantages of both constructions.


\end{document}